\documentclass{svproc}
\usepackage[utf8]{inputenc}

\usepackage{url}

\usepackage[sectionbib]{natbib}
\bibpunct{(}{)}{;}{a}{}{,}%
\usepackage{graphicx}
\usepackage{amsmath,amssymb}
\usepackage{multirow}
\usepackage{longtable}

\usepackage[b5paper]{geometry}
\begin{document}
	\mainmatter              
	\title{Observations of ultraluminous X-ray sources with the 6-m telescope of the SAO RAS}
	\titlerunning{Observations of ultraluminous X-ray sources}  
	%
	\author{A. Vinokurov\inst{1} \and K. Atapin\inst{2} \and
		Y. Solovyeva\inst{1} \and A. Sarkisyan\inst{1} \and D. Oparin\inst{1}}
	\authorrunning{A. Vinokurov et al.} 
	%
	%
	\institute{Special Astrophysical Observatory, Nizhnij Arkhyz, 369167, Russia,\\
		\and
		Sternberg Astronomical Institute, Lomonosov Moscow State University, Universitetskij Pr. 13, Moscow 119992, Russia}
	
	\maketitle

\begin{abstract}
The nature of ultraluminous X-ray sources (ULXs) is not yet fully understood but it can be done in the most efficient way by analyzing its optical emission and environment. We present preliminary results of photometric and spectral studies of some newly identified ULX optical counterparts. Based on the X-ray catalog of ULX candidates by \cite{Liu2011} we have unequivocally identified in the optical range more than 100 sources, for 49 of them we carried out a long slit spectroscopy. According to these results, the majority of the most brightest candidates (m$_V < 22^m$) turned out to be background objects of different types, but the observed parameters of other sources of the sample allow us to classify them as real ULX. Here we present results for ten selected sources.
\end{abstract}

\paragraph{\bf Introduction.}

Ultraluminous X-ray sources are  binary systems with peak X-ray luminosities exceeding the Eddington limit for stellar-mass black holes ($> 10^{39}$ erg\,s$^{-1}$, \citealt{Kaaret2017}) and observed in galaxies at distances up to $\sim100$ Mpc. Such high luminosities can be explained by supercritical accretion onto neutron stars and stellar mass black holes, or standard accretion onto intermediate-mass black holes. Optical studies provide an important component of our knowledge about these unique objects because it allows to determine type of the donors \citep{Liu2013,Motch2014} and investigate parameters of gas outflowing from the accretion disks \citep{Fabrika2015,Kostenkov2020}. These data are extremely important for testing the population synthesis models of ULXs \citep{Wiktorowicz2017}.

We are conducting a multi-wavelength study of the sources from the \cite{Liu2011} catalog containing 479 ULX candidate in 188 galaxies. This paper reports preliminary results of this research, the optical spectroscopy of ULX candidates is still in progress and can potentially change our final conclusions. Because of limited space, in this paper we provide spectroscopic results only for ten sources.

\vspace{-0.3 cm}
\paragraph{\bf Results.}

To identify the optical counterparts we used images from the Chandra X-Ray Observatory and the Hubble Space Telescope (HST). The overall 90\% uncertainty circle of Chandra absolute positioning has a radius of 0.8$^{\prime\prime}$. To improve relative astrometry between the Chandra and HST we used reference sources clearly seen in both ranges (quasars in most cases). If such sources were absent in the HST field of view (FOV), we additionally involved SDSS data. If a reference source appeared to be the only one within the FOV, we took its spectrum to determine its type and prove its ability to radiate in X-rays.

The HST data is available for more than 300 sources of the Liu catalog. For most of them ($\gtrsim$60\%) the identification is ambiguous. The unique identification is obtained for a little more than 100 sources, and 24 of them turned out to be published supernovae, active galactic nuclei (AGN) or objects of our Galaxy. About the same number of sources with high optical to X-ray flux ratios are previously unknown objects presumably of the same types. Some ULX candidates in galaxies of early types coincide with relatively bright sources, whose magnitudes and colors correspond to those of globular clusters. 

Thus we conclude that majority of candidates with bright counterparts (m$_V < 22^m$) actually are not ULXs. We confirmed the nature of five of them spectroscopically (see examples below)  using SCORPIO focal reducer of the Russian 6-m telescope \citep{AfanasievMoiseev2005}. The long slit spectra of another 3 relatively bright ($\approx22^m$) sources appeared to be too noisy for reliable classification. For 41 weak (m$_V > 23^m$) counterparts we carried out spectroscopy of their environment (nebulae and/or star clusters). We found that the absolute magnitudes M$_V$ of these sources are in the range $-3^m \div -7^m$ with rare exceptions like NGC\,3628 X-1 (see below), which is in good agreement with the previously studied ULXs \citep{Vinokurov2018}.

\vspace{-0.3 cm}
\paragraph{\bf Study of selected sources.}

Optical spectroscopy was carried out with the 1200B (spectral range of $3600-5400$\AA\AA, resolution of 5.5\AA), 1200G ($4000-5700$\AA\AA, 5.2\AA) and 1200R ($5700-7500$\AA\AA, 5.2\AA) grisms. The data reduction was done with the LONG context in MIDAS using standard algorithm.

Five out of ten selected objects look normal for ULX counterparts: they have typical for ULX absolute magnitudes M$_V$ from $-4.07\pm0.12$ (NGC\,5585 X-1) to $-6.94\pm0.04$ (NGC\,7714 X-3); all of them are located not far from star formation regions. M$_V$ was estimated from aperture photometry on HST data using APPHOT (IRAF) after correction for reddening. The sixth source, NGC\,3628 X-1, has an abnormally high brightness of M$_V=-9.7\pm1.0$; however, it is surrounded by dust and its reddening estimate is not very reliable. 

In the cases of NGC\,3310 X-3, NGC\,3628 X-1, NGC\,5194 X-2 and NGC\,5585 X-1 we obtained spectra of nebulae located near these ULXs. Using observed ratios of the hydrogen lines in the nebulae and assuming Case B of photoionization we estimated the reddening values A$_V$ as $0.8\pm0.2, 3.2\pm1.0, 0.9\pm0.3, 0.18\pm0.10$, respectively. The nebulae around NGC\,3310 X-3 and NGC\,5194 X-2 are more likely associated with star-forming regions rather than with the ULXs themselves. In contrast to them, the NGC\,5585 X-1 nebula may be a jet-powered ULX bubble (the existence of this nebula was mentioned for the first time by Roberto Soria, HST Proposal 14060). The nebula around NGC\,3628 X-1 may also be excited by ULX. In the cases of NGC7714 X-2,3 we did not detect any nebulae lines and adopted Galactic value of reddening.

\vspace{-0.3 cm}
\begin{figure*}[h!]
	\begin{center}
		\includegraphics[angle=0,scale=0.4]{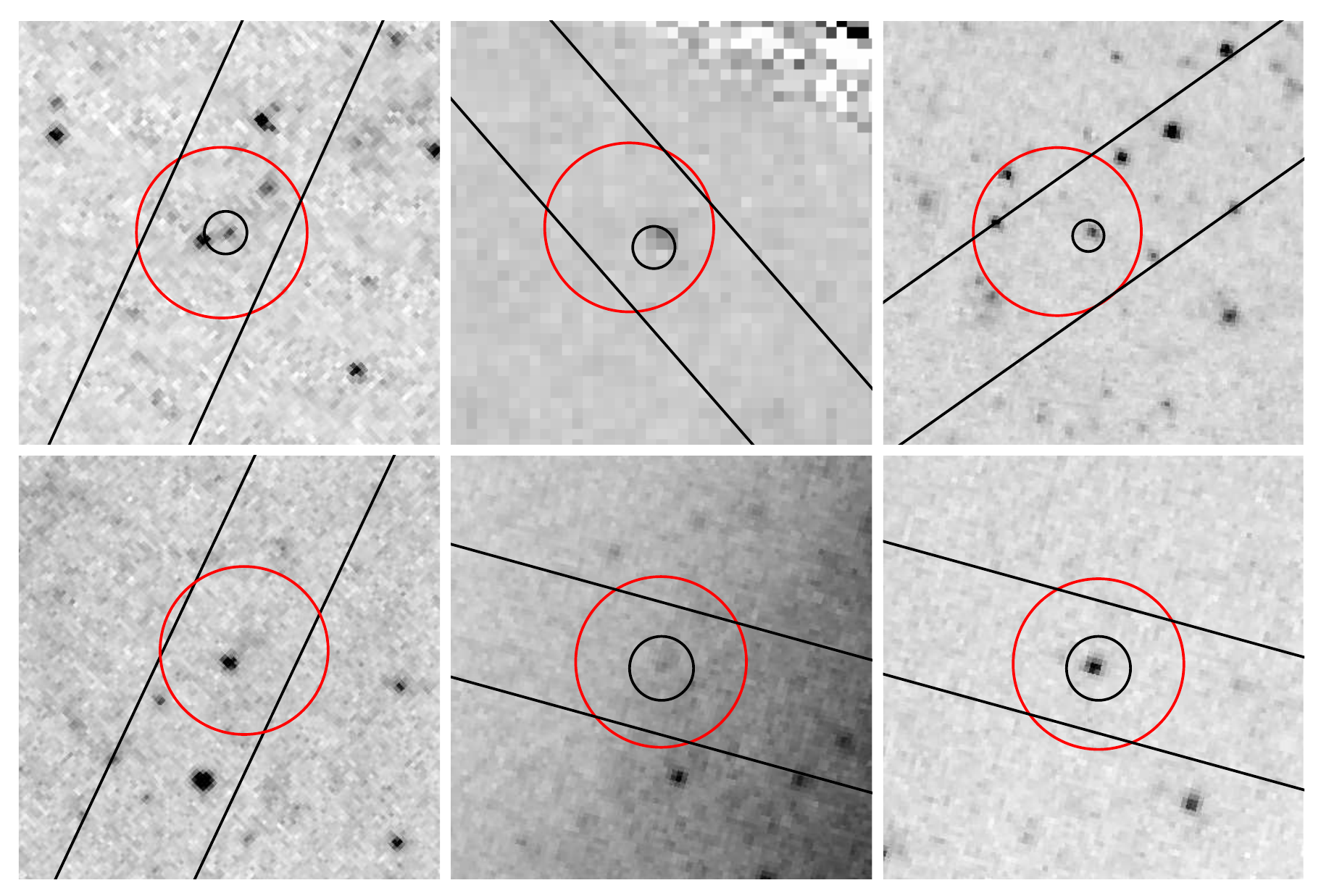}
		\caption{HST images of 4$^{\prime\prime}$ area around the ULXs. Top from left to right: NGC\,3310 X-3 (ACS/F625W), NGC\,3628 X-1 (WFPC2/F606W), NGC\,5194 X-2 (WFC3/F555W); bottom from left to right: NGC\,5585 X-1 (WFC3/F555W), NGC\,7714 X-2 and NGC\,7714 X-3 (ACS/F606W). Big red (0.8$^{\prime\prime}$) and small black circles indicate the original and corrected Chandra position errors of the ULXs. The lines show the position of the 1.2$^{\prime\prime}$ width slit during spectroscopy, north is at the top.}
	\end{center}
	\label{fig:fig1}
\end{figure*}
\vspace{-0.3 cm}

NGC\,628 X-7 turned out to be a Galactic M star with G-band magnitude of 13.68. The source is projected onto a crowded region of the galaxy, but the reliability of the optical identification is beyond doubt, since the proper motion measured on Chandra data ($0.095^{\prime\prime}\pm0.010^{\prime\prime}\,yr^{-1}$) coincides with the value from Gaia DR2 catalog. NGC\,2782 X-7 and NGC\,7541 X-2 are AGNs at redshift of 1.07 and 1.38, respectively. NGC\,5953 X-3 is also a candidate for AGN, but we have only its red spectrum, which makes it difficult to determine the redshift. Possible values are 1.37, 2.49 and 3.30.

\begin{figure*}[h!]
	\begin{center}
		\includegraphics[angle=0,scale=0.35]{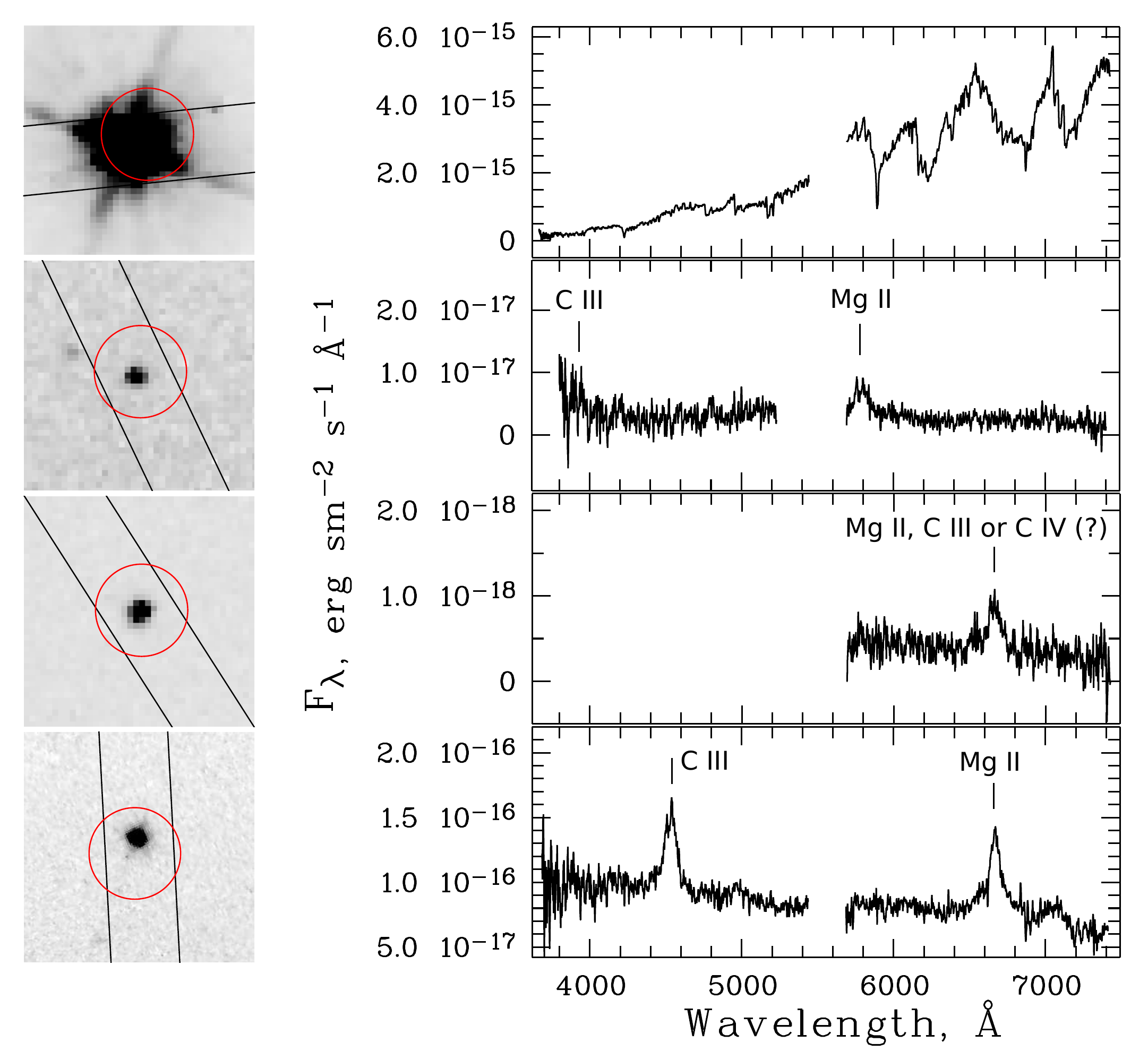}
		\caption{HST images and spectra of NGC\,628 X-7 (WFPC2/F606W), NGC\,2782 X-7 (WFPC2/F555W), NGC\,5953 X-3 (WFPC2/F606W) and NGC\,7541 X-2 (WFC3/F555W; from top to bottom). The designations are the same as in Fig.~\ref{fig:fig1}. The last three spectra are smoothed to the detector spectral resolution of 5.5\AA. The lines were identified using a template QSO spectrum from the Runz code.}
	\end{center}
	\label{fig:fig2}
\end{figure*}
\vspace{-0.3 cm}

\paragraph{\bf Acknowledgements.}
The study was funded by RFBR according to the research project 18-32-20214.

\vspace{-0.3 cm}
\bibliographystyle{aa}
\bibliography{vinokurov.bib}
\end{document}